\documentclass{eas}
\usepackage{graphicx}
\usepackage{amsmath,amssymb}
\usepackage{eurosym}

\TitreGlobal{An astronomical observatory at Dome C, Antarctica for the next decade}
\begin{document}

\title{Development of the opto-mechanical design for ICE-T}
\runningtitle{Di Varano \etal, ICE-T Opto-mechanical design status}
\author {I. Di Varano} \address {Astrophysical Institute Potsdam (AIP), An der Sternwarte 16, 14482 Potsdam, Germany}
\author {K. G. Strassmeier} \sameaddress{1}
\author{T. Granzer} \sameaddress{1}
\author{M. Woche} \sameaddress{1}

\begin{abstract}
ICE-T (International Concordia Explorer Telescope) is a double 60 cm f/1.1 photometric robotic telescope, on a parallactic mount, which will operate at Dome C, in the long Antarctic night, aiming to investigate exoplanets and activity of the hosting stars. Antarctic Plateau site is well known to be one of the best in the world for observations because of sky transparency in all wavelengths and low scintillation noise. Due to the extremely harsh environmental conditions (the lowest average temperature is -80$^\circ$C) the criteria adopted for an optimal design are really challenging. Here we present the strategies we have adopted so far to fulfill the mechanical and optical requirements.
\end{abstract}
\maketitle
\section{Introduction}
ICE-T is a consortium of several European institutes: AIP and AWI from Germany; University of Padua, INAF Observatory of Catania, University of Perugia, ISAC-CNR in Bologna, from Italy; IEEC in Barcelona, Spain. The project has passed through the preliminary design phase in November 2008, where the development of several work packages has been widely discussed. Details concerning optical design, the tube and mount mechanical design, with the outcomes from stress and deformation analysis, control system strategy, data analysis and storage have been defined and approved. 
We intend  to proceed now to the final stage, asking for quotations of optical components and defining manufacturing phases.

High photometric precision will be achievable taking advantage of two main properties of the site. The first is the long winter night which permits precise monitoring of more than 100,000 stars (Strassmeier {\em et al.\/} \cite{st2008}). The second is low scintillation noise which makes the performance of a 60 cm telescope equivalent to a 1.5 m telescope located in a temperate site. Anyway the photon noise sets the limit magnitude to 8th magnitude, whereas in thermal site can be 3 times fainter (Strassmeier \cite{st2007}). Therefore it's necessary to defocus in order to observe targets in the range of 9-19 mag in V.
Primary envisioned scientific targets are extra-solar planets as well as stellar magnetic activity and non-radial pulsations in the structure of the host star. Long term observations with high precision are necessary to detect planets and to analyze starspots and flares. We need two separate bandpasses operating simultaneously in order to discriminate a transit from  spot and plages associated to rotation (Carpano {\em et al.\/} \cite{car2003}). Optical afterglows of gamma-ray bursts and micro-lensing effects are also foreseen among the required targets. 
\section{Wide field photometry}  
\subsection{Optimal field selection}
ICE-T will stare at one single field for each campaign. Once it is pointed, the telescope is supposed to keep on tracking the very same field in a fixed position along right ascension axis for the whole season. No guiding mechanism, image derotator nor corotating dome are required.  Therefore redundancy is the main requirement for a robust and reliable system. Both tubes are equipped with two identical CCD 10.3k$\times$10.3k thinned, back-illuminated, 9$\mu$ pixel size, and an image scale of 2.75$''$ per pixel. The total angular field of view  will be ca. 12.4$^\circ$, equivalent to 8.1$\times$8.1 square degrees. There's no mosaicing, so that astrometric accuracy  across the whole chip will be not affected.
Sloan \textit{g} (402--552 nm) and \textit{i} (691–--818nm) filters are preferable to the Johnson-Cousins as they cut off the emission line of visible aurora (at 5577 \AA). We intend to operate in frame-transfer mode, without shutter, where only the central half is exposed (5300$\times$10600 px), while the rest of it, divided in two quarters, is shaded and used to transfer the charge to the 16 10--MHz amplifiers (8 by side). So the image is partitioned in sixteen parts, each of 2,650$\times$1,325K px.
For the selection of optimal star fields several parameters have been considered: effects of diurnal air mass and refraction variations, solar and lunar interference, interstellar absorption, overexposing of bright stars and ghosts, crowding by background stars, maximization of stars with possible transits, namely dwarf vs giant stars (F\"ugner {\em et al.\/} \cite{fue2009}). Two best fields per operation case have been sorted out investigating 19 different trial regions, taken from NOMAD (Zacharias {\em et al.\/} \cite{zac2005}) and other catalogs. 

With an image scale of 2.75 arcsec/px and an estimated poor seeing of 1.3$''$ (Aristidi {\em et al.\/}\cite{ari09}), PSF would be significantly undersampled. To overcome this issue, defocussing will be employed, a procedure also used by CoRoT (up to 30$''$) and Kepler mission (Christensen-Dalsgaard {\em et al.\/} \cite{chri2007}) for astroseismology. Defocussing ICE-T up to 12 pixels would allow to record 8.5 mag stars with 10 s exposures. The estimated 5$\sigma$ limiting magnitude of the telescope with the envisioned CCDs is V$\cong$20 mag in 10 seconds, reaching a photometric precision of 0.01mag for a V=17mag star. For exposures of 300sec, the limiting magnitude will be 22mag. A limiting magnitude of V=20mag for a 10 second exposure implies that about 100,000 stars brighter than 16.5th magnitude could be detected in a stacked 300 sec exposure (30x10 s frames) in a field at 0-10 degrees from the galactic plane. Of these, about 1000 stars would reach photometric precision as good as 200 $\mu$mag, 30,000 stars would have photometry better or equal to 0.3-1 mmag, another 70,000 stars would have magnitudes precise to roughly 5mmag.
 \subsection{Data storage and analysis} 
Considering a total number of 224 million px for both cameras, one single FITS will have the size of 450 MB. Assuming a 4 months campaign, with 20\% of bad weather, usable time is of 8.3 million seconds. With an integration time of 10 s in frame-transfer mode, the utmost amount of data is $\approx$ 187 TB; combining  30 of the 10-sec frames it decreases to 6.2TB;  data can be furthermode compressed to 3.5 TB. The idea is to store data by 3 levels of redundancy:
\begin{itemize}
	\item combined frames are saved to hard disks that, in case of procurement in 2010, can reach a size of 2-5 TB;
	\item individual frames can be saved into 512--GB SDLT tape robots;	\item one-dimensional light curves from the on-line data reduction, with a required space of $\approx$100 GB, will be saved on a 176 GB FFD (fast flash disk). The optional satellite up- and downlink is not feasible at the moment:  transfer of raw frames in real time,  at a rate of 14MB/s, would require a bandwidth of roughly 140Mbps.
\end{itemize}
In order to determine rotational periods two algorithms will be used independently, within  ARCO reduction package (Di Stefano {\em et al.\/} \cite{distef07}), one based on Scargle-Press period detection routine (Scargle \cite{sca82}), the other on CLEAN, developed by Roberts (\cite{rob87}). 
\section{Adopted design criteria}
\subsection{Optical design}
\begin{figure}
\centering
	\includegraphics[width=12.5cm]{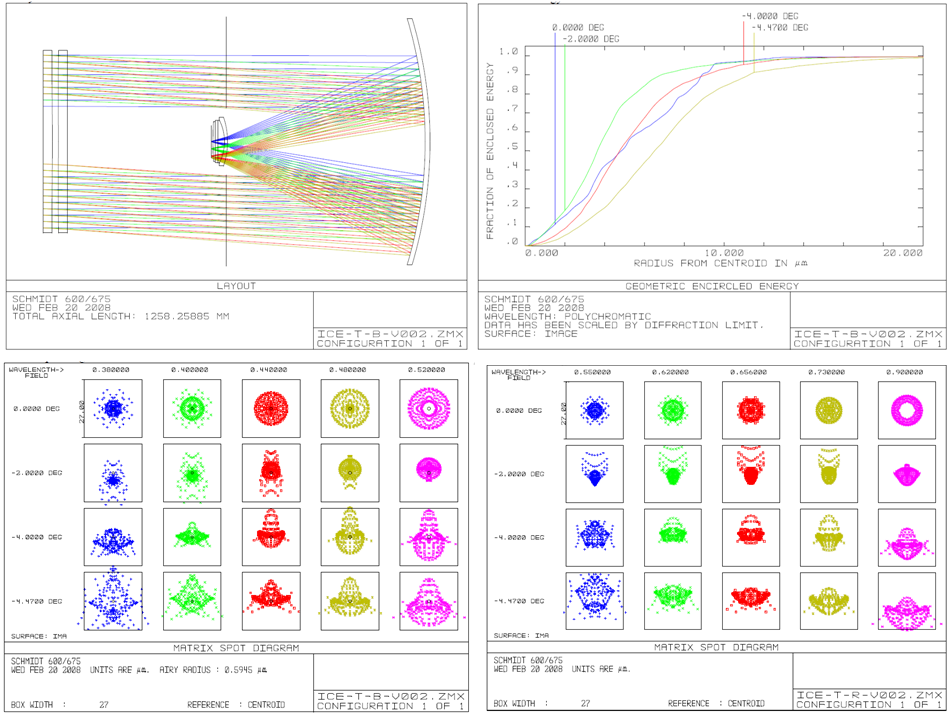}
	\caption{On top left subfigure is shown the Zemax drawing of the f/1.1 Schmidt with two corrector plates, and a 20mm gap; top right represents
the encircled energy fraction in Sloan \textit{g} at different distances from the optical axis in $\mu$; bottom left  and bottom right picture contain respectively the spot diagrams in Sloan \textit{g} and Sloan \textit{i}.}
	\label{figura1}
\end{figure}

The optical system is based on an achromatic Schmidt plate, very fast f/1.125. Main mirror is spherical, made of Zerodur, with a diameter of 820 mm, and radius of curvature of 1414.5 mm. Two corrector plates per tube, of 61 cm diameter are made of different glasses, N-BK7 and N-F2, selected from Schott catalog. An achromat of $\O$180 mm  leads to a flat focal plane. Field--flattener lenses have spherical surfaces and are made of BK7 and have slightly different parameters for Sloan \textit{g} and Sloan \textit{i} filters. In Fig.\ref{figura1} spot diagrams are shown in the two bandpasses. Considering the box is 27$\mu$ wide, best diameter value is of 9$\mu$ at 400 nm.
\subsection{The tube structure}
The  tubes are completely covered by an outside cladding – except the front corrector plate of the optical subsystem. The cladding is formed in a very smooth way avoiding sharp edges or corners, in order to prevent ice accumulation. The overall dimensions of the cladding cylinder are 1520 mm in height, and 1200 mm in width. The tube structure can be divided in a few subsystems: tube cylinder with the mount interface; the front ring on which the corrector holder is mounted; primary mirror cell; 3 spiders and camera interface.

\begin{figure}
\centering 
	\includegraphics[width=14.5cm]{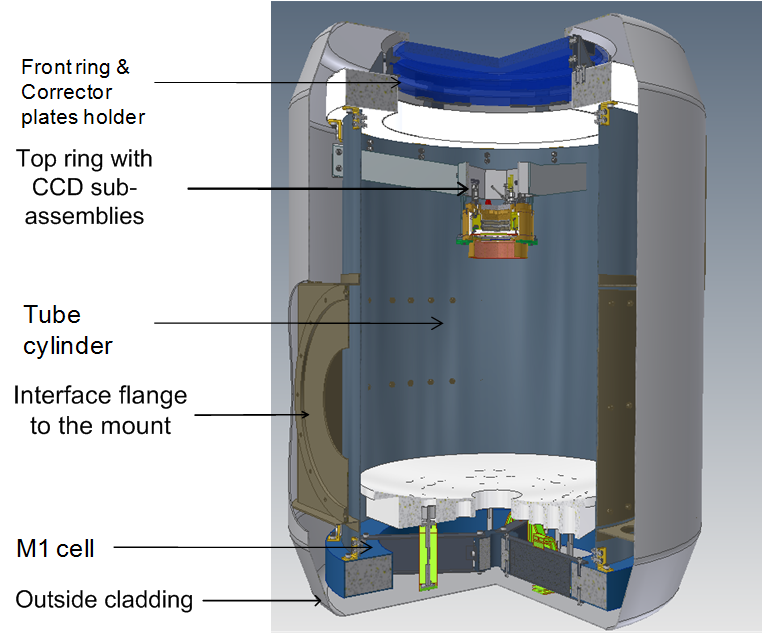}
	\caption{3D section view of the tube with the main assemblies.}
	\label{figura2}
\end{figure}

The cladding consists of smooth glass fiber shells with a wall thickness of 3mm and an outer smooth aluminized surface, e.g. Mylar foil. A light weight Zerodur mirror has been proposed, with holes drilled from the rear. This improves also the thermal behavior, besides
giving advantages in transportation. The corrector plates are connected with the front ring of the tube structure via six Invar brackets and related Titanium flexures. The gap between the corrector plates is used for air ventilation and related heating. The primary mirror is supported by  passive hindle type mounts, with 6 iso-static support points, distributed on three whiffle tree branches. The axial rods are linked to the mirror by Invar pads glued into corresponding holes of the mirror. The lateral supports are attached to the cell structure via Invar brackets as well, and similar flexure arms of titanium are employed (see fig.\ref{figura2}). The parts of the tube structure are made of carbon fiber compound, consisting of five glued layers with different orientation and  mechanical properties (i.e. Young Modulus) optimized to have in principle a coefficient of thermal expansion very close to zero. Outcomes from the static analysis of the all tube structure have revealed a maximum stress reached in horizontal position of $\sigma$=49 MPa. Different heating possibilities are envisaged in order to avoid frost formation (Durand {\em et al.\/} \cite{dur08}) on the outer surface of N--BK7 plate: either with external or internal heating. In the first case infrared lamps, installed inside the dome, can be used, or heaters can be placed on the cladding at the outer rim of the front plate. In the latter case heat is transfered by natural convection to the front surface of the front plate. The advantage of this system is that it can be easily applied on the cladding without the need of outside brackets. The best solution might be that of using the heat produced internally by the camera electronics, and transporting it by an airflow in the gap between the outer and inner corrector plate. CFD calculations of different models  are being investigated right now to check the best solution.
\subsection{The mount}
The mount frame has been designed optimizing nine geometric parameters in order to have  at the same time a light structure with low deformations. It consists of welded metal plates of forged carbon steel ASTM A350 LF2 of 10mm thickness. Overall size are 450$\times$840$\times$990 mm. On top of the mount frame the declination assembly is mounted, which includes: the hollow shaft with a maximum external diameter 381 mm, internal one of 280 mm; two taper roller bearings, (type 32972 and L865547 by SKF), mounted in O configuration so that they can compensate for thermal variation, keeping a precise alignment; hollow magnetic encoder provided by Heidenhain, ERM 452 $\O$410, suitable with the shaft dimension; declination motion is achieved by means of a gear sector with a 3.5 module.
We are considering to use the same type of drives on both right ascension and declination axes, in order to have a modular system. To avoid any backlash problem double pinions rotating in opposite direction are foreseen, controlled via an inverter (Tosti {\em et al.\/} \cite{tosti06}). We have  contacted so far the company REDEX-ANDANTEX, and we are choosing a Dual Drive typology, with a torsional joint mechanically preloaded.
Same stepper motors by Pythron are envisioned both in RA and dec, in particular of ZSH 87/2 class, with 2 phases, in agreement with  the estimated value of the torque moment applied on the motor shaft, Cm=1.7 Nm. It has an accuracy of 1.8$^\circ$/s, with optional incremental  encoders H500 with up to 500 lines of resolution.
\section{Conclusions}
The telescope is mounted via an adjustable platform on three concrete piers of $\O$600mm, distributed on an outer circle of 1.8m diameter. The instrument is hosted by a dome produced by Baader Planetarium, with an inner diameter of 3.4m. The dome departed from Bremerhaven in December 2007 and was successfully installed at Dome C at the beginning of 2009. The TAVERN  Star photometer  will be hopefully shipped to Dome C in September 2009, so that it will start the winter campaign in February 2010.

\end{document}